\documentclass[aps,pra,10pt,superscriptaddress,twocolumn]{revtex4}

\usepackage{amsmath}
\usepackage{graphicx}
\usepackage{placeins}
\usepackage{times}
\usepackage{amssymb}
\usepackage{hyperref}
\usepackage{xcolor}
\usepackage[T1]{fontenc}
\usepackage{bm}

\begin{document}
\title{Casimir--Polder induced Rydberg macrodimers}
\author{Johannes Block}
\email{johannes.block@uni-rostock.de}
\author{Stefan Scheel}
\email{stefan.scheel@uni-rostock.de}
\affiliation{Institut f\"ur Physik, Universit\"at Rostock, 
Albert-Einstein-Strasse 23-24, D-18059 Rostock, Germany}
\date{\today} 

\begin{abstract}
We theoretically investigate Rydberg atom pair potentials of Rb atoms in front 
of a perfectly conducting plate. The pair potentials are perturbed by both the 
Casimir--Polder potential acting on a single atom and the scattering contribution 
to the interatomic interaction. In contrast to the pair potentials in free 
space, at atom-surface distances $d_s \lesssim 4\,\mu$m, avoided crossings 
appear. In the associated potential wells that are entirely due to dispersion 
interactions with the surface, there exist vibrational bound states, i.e. 
Rydberg macrodimers, with groundstate energies of up to $E = -68\,$MHz and 
radial expectation values of the order of several $\mu$m.
\end{abstract}

\maketitle

\section{Introduction}
Rydberg atoms are known for their exaggerated properties due to their large 
wavefunction extension and atomic dipole moments \cite{Gallagher}. For example, 
interatomic van der Waals potentials scale with the principal quantum number 
$n$ as $n^{11}$. Similarly, dispersion interactions between a single atom and a 
macroscopic body, referred to as Casimir--Polder potentials 
\cite{Scheel2008,Buhmann2012}, typically show a scaling with $n^4$ 
\cite{Crosse2010}, making Rydberg atoms highly susceptible to changes in their 
immediate environment. This is equally true for externally applied or, indeed, 
stray electric fields due to adsorbates 
\cite{Tauschinsky2010,Hattermann2012,Epple2014,Langbecker2017}.

Interactions of Rydberg atoms with surrounding ground-state atoms or other 
Rydberg atoms provide the basis for novel classes of molecules.
Low-energy scattering between a Rydberg electron and a ground-state atom leads 
to the formation of long-range Rydberg molecules  
\cite{Greene2000,Bendkowsky2009,Butscher2011,Saßmannshausen2016a,Shaffer2018}.
Their wave function can form distinct shapes like the famous trilobite molecule \cite{Greene2000}.
Rydberg macrodimers between two Rydberg atoms, on the other hand, are the 
result of strong multipolar or van der Waals interatomic interactions with or 
without background electric fields 
\cite{Boisseau2002,Farooqi2003,Schwettmann2007,Samboy2011,Samboy2011a,
Saßmannshausen2016}.
Due to the complex level structure of Rydberg atoms, many-body bound Rydberg 
states exist as well \cite{Kiffner2013add,Kiffner2014a}.

In this work, we show that the presence of a (conducting) surface leads to 
the formation of macrodimers without external background fields being present.
Indeed, the surface-induced image multipoles in a sense replace an external 
electric field. At interatomic distances and atom-surface distances in the 
range of a few $\mu$m, we find that avoided crossings appear in the energy 
spectrum of a pair of atoms, leading to potential wells that support a large 
number of vibrational bound states.

This article is organized as follows. In Sec.~\ref{sec:setup} we set the stage 
for calculating pair potentials between Rydberg atoms near conducting surfaces 
by reviewing how dispersion forces near surfaces alter the level structure of 
atoms as well as their interatomic interaction potential. The results of our 
calculations of the pair potentials containing bound vibrational states are 
presented in Sec.~\ref{sec:results}, with a discussion and conclusions being 
provided in Sec.~\ref{sec:conclusions}.

\section{Rydberg atoms near surfaces}
\label{sec:setup}

\begin{figure}
 \includegraphics[width=\columnwidth]{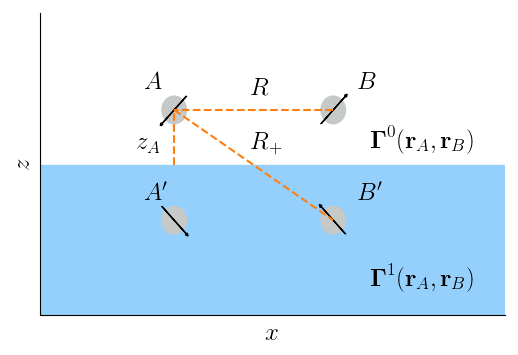}
 \caption{Two atoms $A$ and $B$ in free space in front of a perfectly 
conducting plate (blue). The atoms are located in the half space with $z>0$ at 
a distance $z_A = z_B = d_s$ away from the surface with an interatomic 
separation $R$. Image dipoles are created at $z' = -z_B = -z_A$. The distance 
between one atom and the image of the second atom is $R_+$. 
The Green tensors $\boldsymbol{\Gamma}^0$ and $\boldsymbol{\Gamma}^1$ represent 
the free-space and scattering contribution of the interaction, respectively.}
 \label{fig:atoms_with_surface}
\end{figure}

We begin by detailing the envisaged scenario in which two $^{87}$Rb atoms in 
high-lying Rydberg states are held in free space in close proximity to a 
(perfectly) conducting half space as shown in Fig.~\ref{fig:atoms_with_surface}. 
Without loss of generality, but rather to aid simplicity, we assume that both 
atoms are held at the same distance $d_s$ to the surface.

Their dynamics is governed by the Hamiltonian
\begin{align}
  \hat{H} =  \hat{H}_A + \hat{H}_B + \hat{H}_\text{int}
\end{align}
which consists of the single-atom Hamiltonians $\hat{H}_i$ for atom $i$ and the 
interaction Hamiltonian $\hat{H}_\text{int}$. Let us first consider the $\hat{H}_i$ 
for atoms in free space which contain the unperturbed single-atom 
Rydberg energy levels of each atom that can be efficiently computed, e.g. by 
the \textsc{pairinteraction} software \cite{Weber2017}. 
Rydberg atomic energies can be written as
\begin{equation}
  E_{nlj} = -\frac{hcR^\ast}{(n-\delta_{nlj})^2}
  \label{eq:rydberg}
\end{equation}
with the modified Rydberg constant $R^\ast$, the principal quantum number $n$ 
and the (phenomenological) quantum defect $\delta_{nlj}$ which is a function of 
the quantum numbers $n,l,j$.

The Hamiltonian $\hat{H}_\text{int}$ governs the (surface-mediated) interaction 
between the atoms, and will be expanded in terms of the first multipole moments.
The multipole interaction can be expressed using the Green tensor approach of 
macroscopic quantum electrodynamics \cite{Scheel2008}. The Green tensor 
$\mathbf{G}(\mathbf{r}_A,\mathbf{r}_B,\omega)$ can be interpreted as the 
propagator function of a monochromatic electromagnetic wave with frequency 
$\omega$ from a source point $\mathbf{r}_B$ to an observation point 
$\mathbf{r}_A$.

In the presence of a macroscopic body, because of the linearity of Maxwell's 
equations, the Green tensor $\mathbf{G}$ can be decomposed into a sum
\begin{equation}
 \mathbf{G}(\mathbf{r}_A,\mathbf{r}_B,\omega) = \mathbf{G}^0(\mathbf{r}_A,\mathbf{r}_B,\omega) + \mathbf{G}^1(\mathbf{r}_A,\mathbf{r}_B,\omega)
 \label{eq:greentensor}
\end{equation}
comprising of the free-space tensor $\mathbf{G}^0$ and a scattering 
contribution $\mathbf{G}^1$. The Green tensor in Eq.~(\ref{eq:greentensor}) can 
be simplified significantly in the nonretarded limit. Due to the small energy 
spacing of adjacent energy levels in Rydberg atoms, one can safely assume that 
all distances (interatomic as well as atom-surface distances) obey $d \ll 
c/\omega_\text{max}$ with the maximum of all relevant transition frequencies 
$\omega_\text{max}$. In this regime, we may take the static (frequency-independent) 
limit of the Green tensor $\mathbf{\Gamma}(\mathbf{r}_A,\mathbf{r}_B) = 
\lim_{\omega \to 
0}\frac{\omega^2}{c^2}\mathbf{G}(\mathbf{r}_A,\mathbf{r}_B,\omega)$.

The interaction Hamiltonian can then be written as
\begin{equation}
 \begin{aligned}
  &\hat{H}_\text{int} = \frac{1}{\varepsilon_0}\left[\hat{\mathbf{d}}_A\cdot\mathbf{\Gamma}(\mathbf{r}_A,\mathbf{r}_B)\cdot\hat{\mathbf{d}}_B\right.\\
  &+\left. \hat{\mathbf{d}}_A\cdot\mathbf{\Gamma}(\mathbf{r}_A,\mathbf{r}_B)\overleftarrow{\nabla}:\hat{\mathbf{Q}}_B + \hat{\mathbf{Q}}_A:\overrightarrow{\nabla}\mathbf{\Gamma}(\mathbf{r}_A,\mathbf{r}_B)\cdot\hat{\mathbf{d}}_B\right]
 \label{eq:interaction}
\end{aligned}
\end{equation}
for interactions up to dipole-quadrupole type \cite{Crosse2010,Block2017} with 
the dipole transition operators $\hat{\mathbf{d}}_i$ and the quadrupole 
transition operators $\hat{\mathbf{Q}}_i$ associated with atom $i$. Note that, 
when inserting the static limit of the free-space Green tensor into the first 
term in Eq.~(\ref{eq:interaction}), one recovers the static dipole-dipole 
interaction commonly used in Rydberg physics \cite{Walker2008,Block2017}. 
However, due to the large spatial extent of the electronic wave function of 
Rydberg atoms, it is often not sufficient to only consider dipole-dipole 
interaction which assumes a point-like scatterer, and higher-order multipole 
contributions such as dipole-quadrupole terms have to be considered in addition.

According to the decomposition (\ref{eq:greentensor}) of the Green tensor,
every term in $\hat{H}_\text{int}$ consists of a free-space part and a 
scattering contribution accounting for reflection off the surface.
For two atoms located in the $xz$-plane with the perfect mirror in the
half space $z<0$, the nonretarded Green tensor is given by \cite{Block2017}
\begin{gather}
\bm{\Gamma}(\mathbf{r}_A,\mathbf{r}_B) = -\frac{1}{4\pi} \left[ \frac{1}{R^3}
\begin{pmatrix}
1 & 0 & 0 \\ 0 & 1 & 0 \\ 0 & 0 & 1
\end{pmatrix}
-\frac{3}{R^5}
\begin{pmatrix}
x^2 & 0 & xz_- \\ 0 & 0 & 0 \\ xz_- & 0 & z_-^2
\end{pmatrix}
\right] \nonumber\\
+\frac{1}{4\pi} \left[ 
\frac{1}{R_+^3} 
\begin{pmatrix}
1 & 0 & 0 \\ 0 & 1 & 0 \\ 0 & 0 & 2
\end{pmatrix}
-\frac{3}{R_+^5}
\begin{pmatrix}
x^2 & 0 & -xz_+ \\ 0 & 0 & 0 \\ xz_+ & 0 & x^2
\end{pmatrix}
\right].
\label{eq:staticGmirror}
\end{gather}
Here, we used $x = x_A - x_B$, $z_- = z_A - z_B$, the interatomic distance $R^2 
= x^2+z_-^2$ and  $z_+ = z_A+z_B$ and $R_+^2 = x^2 + z_+^2$ as coordinates in 
the scattering part of the Green tensor. The interpretation of 
Eq.~(\ref{eq:staticGmirror}) is straightforward, with the first line describing 
the direct interaction between the atoms, and the second line their interaction 
with their respective mirror images inside the conducting body. It is the 
latter contribution to the interaction that effectively mimics an external 
electric field, and which gives rise to the state mixing we will encounter soon.

Of course, the dispersion interaction with the surface already affects each
individual atom. This Casimir--Polder interaction can be cast into a similar 
form to Eq.~(\ref{eq:interaction}). For an excited atom at position 
$\mathbf{r}$ in some state $|k\rangle$, the Casimir--Polder potential in 
the nonretarded limit reads \cite{Buhmann2012a}
\begin{equation}
U_k(\mathbf{r}) = 
-\frac{\langle\hat{\mathbf{d}}\cdot
\mathbf{\Gamma}^1(\mathbf{r},\mathbf{r})\cdot
\hat{\mathbf{d}}\rangle_k}{2\varepsilon_0}.
  \label{eq:cppotential_tensor}
\end{equation}
Note that the Casimir--Polder potential (\ref{eq:cppotential_tensor}) 
features only the static scattering Green tensor $\mathbf{\Gamma}^1$.
The free-space contribution associated with $\mathbf{\Gamma}^0$ is the vacuum 
Lamb shift that is already included in the atomic energy spectra according to 
Eq.~(\ref{eq:rydberg}). The scattering contribution 
$\mathbf{\Gamma}^1(\mathbf{r},\mathbf{r})$ in the coincidence limit of its 
spatial arguments is diagonal, and the dipole transition operator can be 
decomposed into components parallel ($\hat{\mathbf{d}}^\|$) and perpendicular 
($\hat{\mathbf{d}}^\perp$) to the surface. The expectation value in the 
numerator of 
Eq.~(\ref{eq:cppotential_tensor}) refers to the atom in state
$|k\rangle_A$ and is given by the sum over all dipole moments
$\langle |\hat{\mathbf{d}}^{\parallel (\perp)}|^2\rangle_k 
= \sum_{k'}|\hat{\mathbf{d}}^{\parallel (\perp)}_{kk'}|^2$.

In principle, following the arguments that led us to include dipole-quadrupole 
interactions in the two-atom potential (\ref{eq:interaction}), one would have 
to include higher-order multipole contributions in the Casimir--Polder 
potential, too. However, angular-momentum selection rules do not allow 
dipole-quadrupole terms in Eq.~(\ref{eq:cppotential_tensor}). The first 
nonvanishing higher-order term contributing to the Casimir--Polder potential 
would be a quadrupole-quadrupole interaction. However, in order to be 
consistent in the truncation at a given multipole order, this would require the 
addition not only of a quadrupole-quadrupole interaction to $\hat{H}_\text{int}$, 
but at the same level of truncation also dipole-octupole (and octupole-dipole) 
contributions, which we take to be excessive.

For an atom in state $|k\rangle$ at a distance $d_s$ from a perfectly 
conducting half space in the nonretarded limit, the Casimir--Polder potential 
is found by inserting Eq.~(\ref{eq:staticGmirror}) with $x=z_-=0$ into 
Eq.~(\ref{eq:cppotential_tensor}) with the result that \cite{Buhmann2012a}
\begin{equation}
 U_k(d_s) = -\frac{\langle 
\hat{\mathbf{d}}^{\parallel2}+2\hat{\mathbf{d}}^{\perp2}\rangle_k}{
64\pi\varepsilon_0 d_s^3}.
 \label{eq:cppotential}
\end{equation}
The full single-atom energy is then the sum of Eqs.~(\ref{eq:rydberg})
and (\ref{eq:cppotential}), 
\begin{equation}
 E_{nlj}(d_s) = E_{nlj} + U_{nlj}(d_s).
\end{equation}

When calculating atomic interactions in free space, the quantization axis is 
often chosen parallel to the molecular axis \cite{Walker2008}.
In this case, the projection of the total angular momentum $M = m_{jA}+m_{jB}$ 
is a conserved quantity reducing the total basis size of interaction 
Hamiltonians considerably \cite{Stanojevic2006,Stanojevic2008}.
However, the presence of an interface breaks the rotational symmetry of the 
problem, and the molecular axis and the normal direction to the surface do not 
necessarily coincide. It is therefore more expedient to choose the 
$x$-axis as the quantization axis \cite{Donaire2015}.
The projection of the total angular momentum $M$ is not conserved in this case and requires a larger 
basis set for the Hamiltonian \cite{Block2017}.

\section{Macrodimers formed by atom-surface interaction}
\label{sec:results}

Our choice of pair states has been informed by their energetic proximity.
A suitable choice of a dipole-coupled set of pair states could be 
$\{|51s_{1/2};53s_{1/2}\rangle, |51p_{1/2};52p_{1/2}\rangle\}$ with 
an energy difference of only $83\,$MHz between the unperturbed states.
In the following, all energies are expressed relative to the 
$|51s_{1/2};53s_{1/2}\rangle$ asymptote for infinite atomic distance at given
surface distance $d_s$. For all pair states mentioned the projection $M$ of the 
total angular momentum is equal to zero, $M = m_{jA} + m_{jB} = 0$.
Their interaction can only be written in multipole form as in
Eq.~(\ref{eq:interaction}) for interatomic distances greater than the LeRoy 
radius $R > R_{LR}=2(\sqrt{\langle r^2\rangle_A}+\sqrt{\langle r^2\rangle_B})$ 
using the 
rms position of the electron of atom $i$. For the $|51p_{1/2};52p_{1/2}\rangle$ 
asymptote, we obtain $R_{LR}\approx 0.80\,\mu$m. As we will later restrict our 
calculations to $R \gtrsim 1.2\,\mu$m, we can safely assume a negligible 
overlap of the electronic wave functions.

\begin{figure}
\includegraphics[width=\columnwidth]{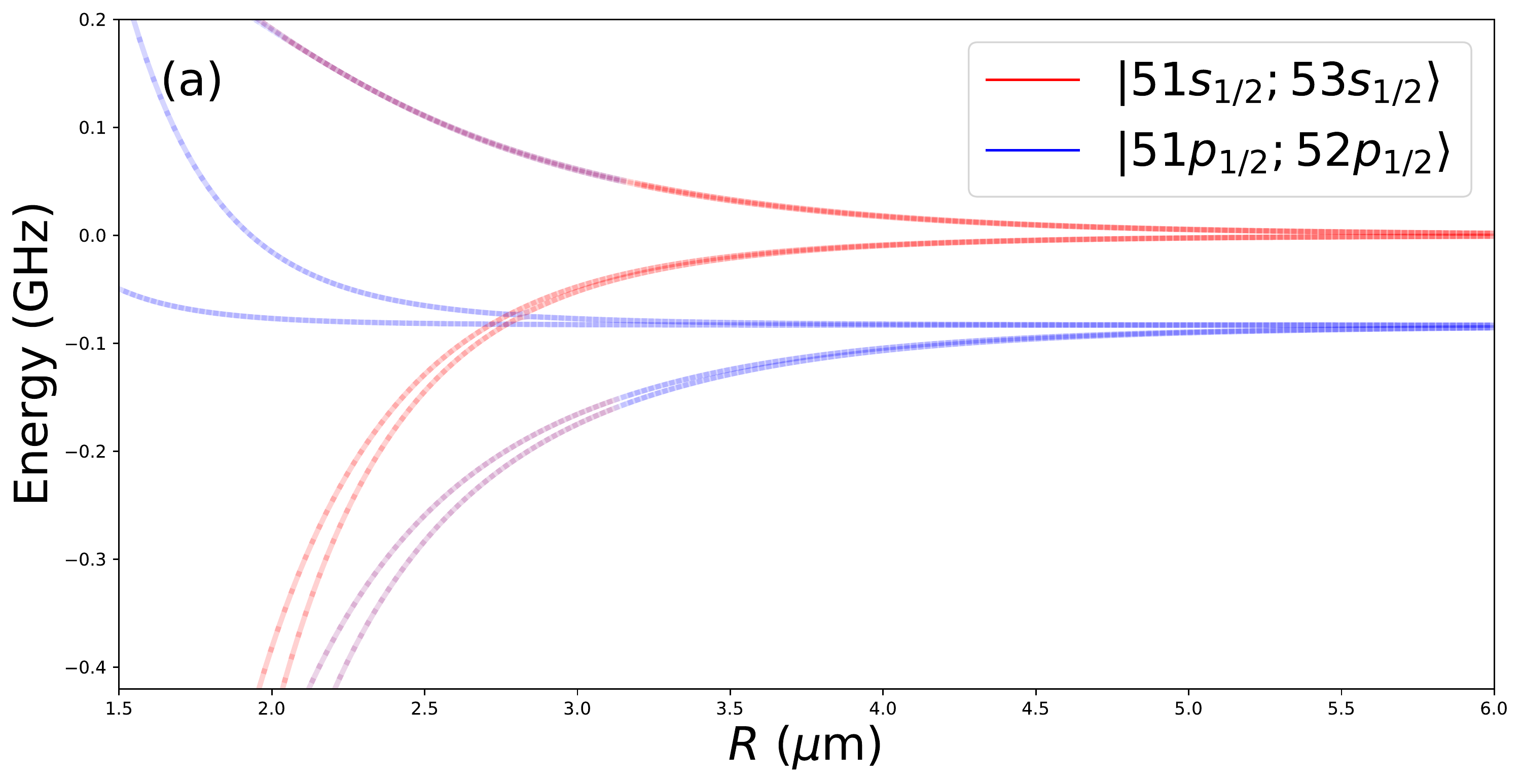}\\
\includegraphics[width=\columnwidth]{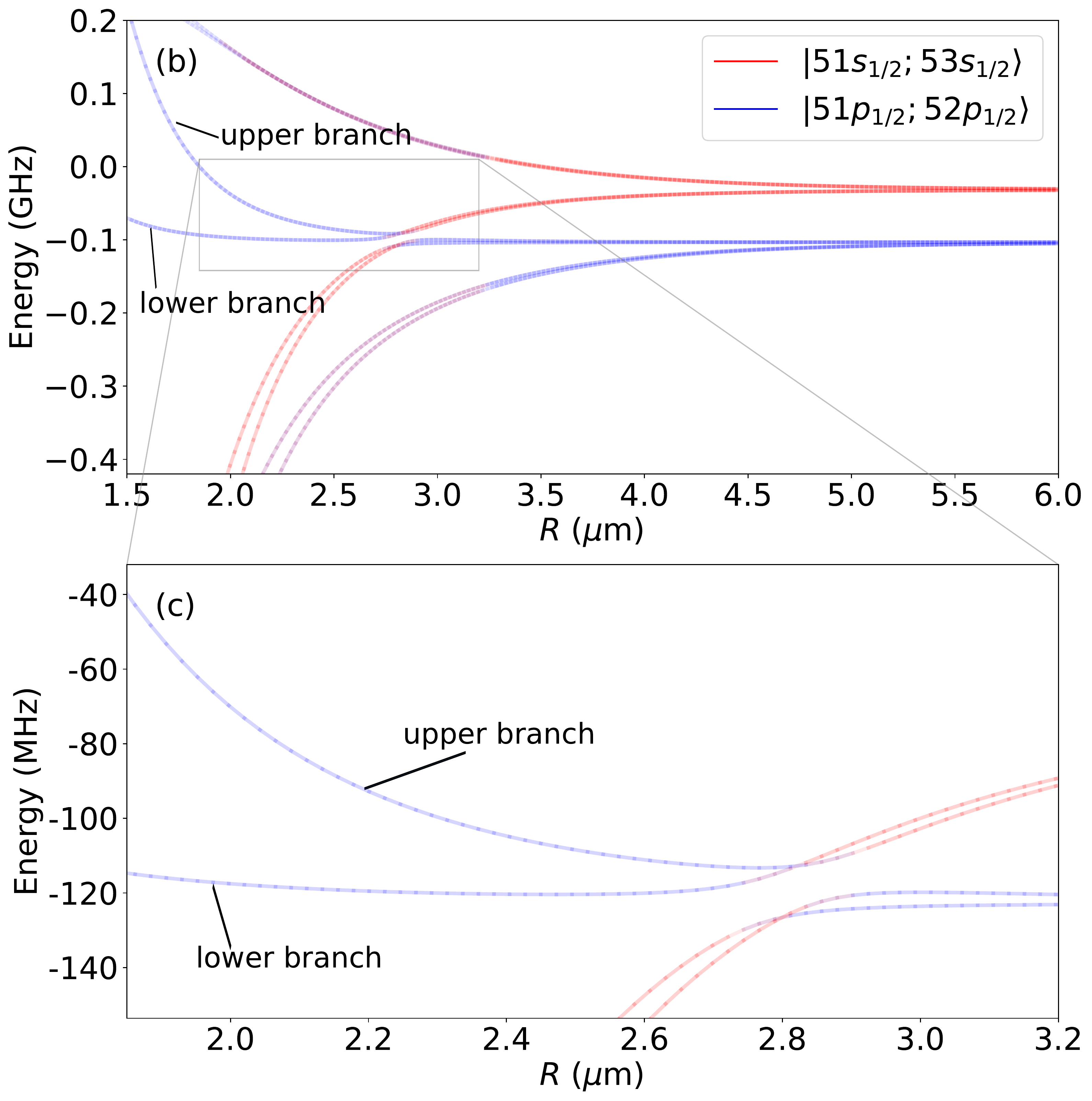}
\caption{Pair potential for Rb Rydberg states around the 
$|51s_{1/2};53s_{1/2}\rangle$ asymptote as a function of interatomic distance 
$R$. Potential curves attributed to the $|51s_{1/2};53s_{1/2}\rangle$ asymptote 
are shown in red, those attributed to $|51p_{1/2};52p_{1/2}\rangle$ states in 
blue. (a) pair potential in free space, (b) potential 
with perfectly conducting plate at $d_s = 3\,\mu$m and (c) detail of 
the avoided crossing region.}
 \label{fig:pairpotential}
\end{figure}

We have added the Hamiltonian (\ref{eq:interaction}) and the potential
(\ref{eq:cppotential}) to the \textsc{pairinteraction} package \cite{Weber2017} 
that numerically computes the spectrum as a function of interatomic separation 
and atom-surface distance. The calculations are conducted for a molecular axis 
that is aligned in parallel with respect to the surface and which are, as has 
been checked numerically, robust under small tilts.
The pair potentials of Rb Rydberg states around the 
$|51s_{1/2}m_j=1/2;53s_{1/2}m_j=-1/2\rangle$ asymptote is shown in 
Fig.~\ref{fig:pairpotential}. In free space, i.e. in the absence of the 
reflecting half-space, the energy levels for the pairs 
$|51s_{1/2};53s_{1/2}\rangle$ and $|51p_{1/2};52p_{1/2}\rangle$ cross 
without disturbance (Fig.~\ref{fig:pairpotential}(a)).

At atom-surface distances $d_s \lesssim 4\mu$m, an avoided crossing 
appears at $R \approx 2.8\,\mu$m between the $|51s_{1/2};53s_{1/2}\rangle$ and 
the $|51p_{1/2};52p_{1/2}\rangle$ asymptotes, and potential wells form.
Figure~\ref{fig:pairpotential}(b) shows the resulting pair 
potential at $d_s=3\mu$m. A detailed investigation of the avoided crossing 
region shows the opening of an energy gap of approximately $13\,$MHz (Fig.~\ref{fig:pairpotential}(c)). 
As the energy gap between the potential curves increases with decreasing atom-surface distance, the potential minimum 
shifts to smaller $R_\text{min}$. We associate these emerging potential wells with 
bound Rydberg-Rydberg states, the so-called Rydberg macrodimers.

\subsection{Rotational vs. electronic timescales}

The upper limits of the lifetimes of these macrodimers is determined by the 
lifetimes of the individual Rydberg states \cite{Schwettmann2007}. 
The potential wells shown for an atom-surface distance $d_s=3\,\mu$m in 
Fig.~\ref{fig:pairpotential} support macrodimer states with vibrational quantum 
numbers $\nu\lesssim 160(228)$ for the upper (lower) branch. 
Dipole-quadrupole interactions as taken into account here would normally give 
rise to rotational-electronic interactions \cite{Deiglmayr2014}. However, as in 
previous studies \cite{Schwettmann2007}, rotational states can be safely ignored 
owing to the timescales involved. 
We can estimate the timescale $\tau$ for rotation of a macrodimer by invoking a 
classical dumbbell model giving $\tau=2\pi\langle R\rangle/(2v)$ with the 
relative velocity $v\approx\sqrt{k_B T/m_{Rb}}$ at temperature $T$. Assuming a 
temperature of the atom cloud of $T=40\,\mu$K and a macrodimer distance of 
$\langle R \rangle = 2.7\,\mu$m, this amounts to $\tau\approx140\,\mu$s which 
has to be compared with the lifetimes of the Rydberg states themselves that are 
also modified by the presence of the surface. With the Casimir--Polder potential 
--- more precisely, the medium-assisted Lamb shift --- and the spontaneous decay 
rate forming a Hilbert transform pair \cite{Scheel2008}, any surface-induced 
level shift is accompanied by a change in the corresponding lifetime. However, 
when computing decay rates, the idealized assumption of a perfectly conducting 
surface can no longer be upheld, and the finite permittivity of the surface has 
to be taken into account.

At sufficiently low temperature at which we can neglect thermal effects 
associated with absorption or stimulated emission, the enhancement of the 
spontaneous decay rate $\Gamma_d$ over its free-space value $\Gamma_0$ is, in 
the nonretarded limit valid here, given by \cite{Buhmann2008}
\begin{equation}
\frac{\Gamma_d}{\Gamma_0}  = \frac{3}{8} \sum_{k<n} 
\left(1+\frac{|d_{nk,z}|^2}{|\mathbf{d}_{nk}|^2}\right) 
\left(\frac{c}{\omega_{nk}d_s}\right)^3
\frac{\mathrm{Im}\,\varepsilon(\omega_{nk})}{|\varepsilon(\omega_{nk})+1|^2}.
\end{equation}
Here, $\mathbf{d}_{nk}$ and $d_{nk,z}$ are the dipole transition moment for the 
$n\to k$ transition and its $z$-component, respectively, and $\omega_{nk}$ is 
its transition frequency. We use a simple Drude model for the permittivity 
$\varepsilon(\omega) = 1 - \frac{\omega_p^2}{\omega^2+i\omega\gamma}$ with 
values for the plasma frequency $\omega_p$ and the damping constant $\gamma$ 
taken from Ref.~\cite{Buhmann2008}. For a gold surface and atomic Rydberg 
states with principal quantum number $n\simeq 50$, we arrive at typical 
enhancement factors $\Gamma_d/\Gamma_0 \approx 10$ at $d_s = 3\,\mu$m. Together 
with the free-space lifetimes calculated for $|50p_{1/2}\rangle$ of $\simeq 
260\,\mu$s \cite{Beterov2009}, we find that $\tau\Gamma_d\gtrsim5$, i.e. a much 
longer rotational timescale than that associated with decay of the Rydberg 
state. 
In addition, there are reports of much shorter lived macrodimer states for 
$62s_{1/2}$ Cs atoms with (vibrational) lifetimes estimated to be $3\ldots 
6\,\mu$s \cite{Han2018}, thus supporting our assertion that the rotational 
spectrum can be safely neglected.

\subsection{Vibrational states}

The main contribution to the potential wells formed by the potential energy 
surfaces shown in Fig.~\ref{fig:pairpotential} shifts from the 
$|51p_{1/2};52p_{1/2}\rangle$  asymptote to the $|51s_{1/2};53s_{1/2}\rangle$ 
asymptote with increasing $R$. There are minor contributions from 
$|50d_{3/2};52s_{1/2}\rangle$ and $|51s_{1/2};53s_{1/2}\rangle$ pair states 
with $M=0$ to the upper potential branch, while the lower branch consists 
almost entirely of the $|51p_{1/2};52p_{1/2}\rangle$ asymptote at the 
small-separation end of the avoided crossing. Due to this state transition 
along the potential energy surface, bound states can only exist in the 
adiabatic limit.

\begin{figure}
\includegraphics[width=\columnwidth]{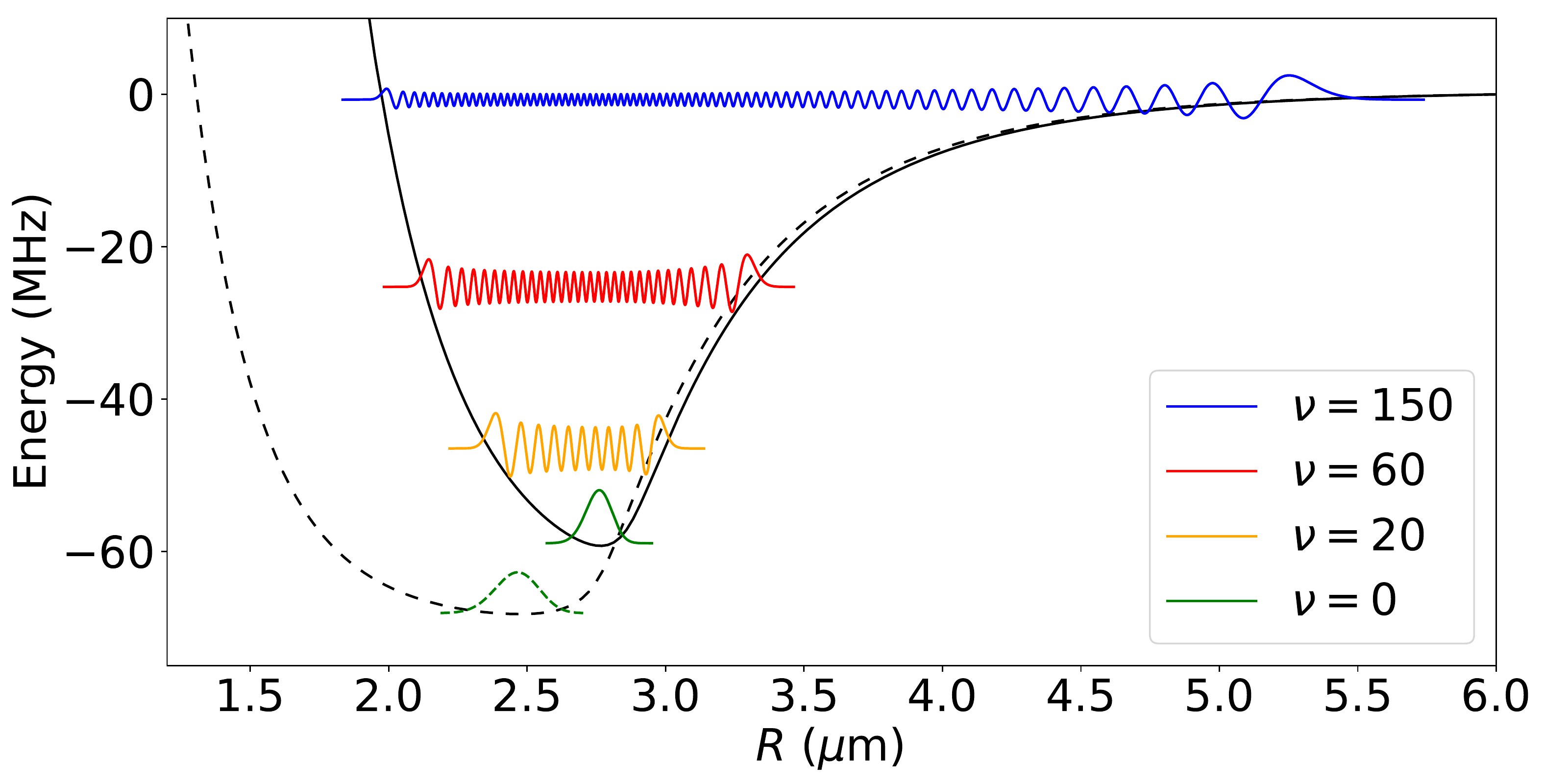}
\caption{Upper (solid) and lower (dashed) potential branches at atom-surface 
distance $d_s = 3\,\mu$m and sketches of vibrational wave functions for $\nu = 
0, 20, 60, 160$ (bottom to top) versus interatomic distance $R$. For the lower 
branch only the ground-state wave function $\nu = 0$ is indicated. The 
expectation value $\langle R \rangle$ is smaller for $\nu = 20$ than for $\nu = 
0$ due to the potential's asymmetry.}
\label{fig:potentials_and_wf}
\end{figure}

The two potential branches associated with the energy surfaces close to the 
avoided crossing (Fig.~\ref{fig:pairpotential}(c)) are depicted in 
Fig.~\ref{fig:potentials_and_wf} including the vibrational wave functions for 
$\nu = 0, 20, 60, 150$ for the upper potential branch. In the potential well 
formed by the lower potential branch, only the ground-state wave function 
$\nu=0$ is shown for comparison. The potential minimum of the lower branch is 
approximately $9$\,MHz lower than that of the upper branch. Using Numerov's 
method \cite{Blatt1967} to find the eigenenergies and wavefunctions of the 
macrodimer states, we find states with a maximal vibrational quantum number 
$\nu_\text{max} \approx 160(228)$ for the upper (lower) branch at $d_s = 
3\,\mu$m. 

\begin{figure}[ht]
\includegraphics[width=\columnwidth]
{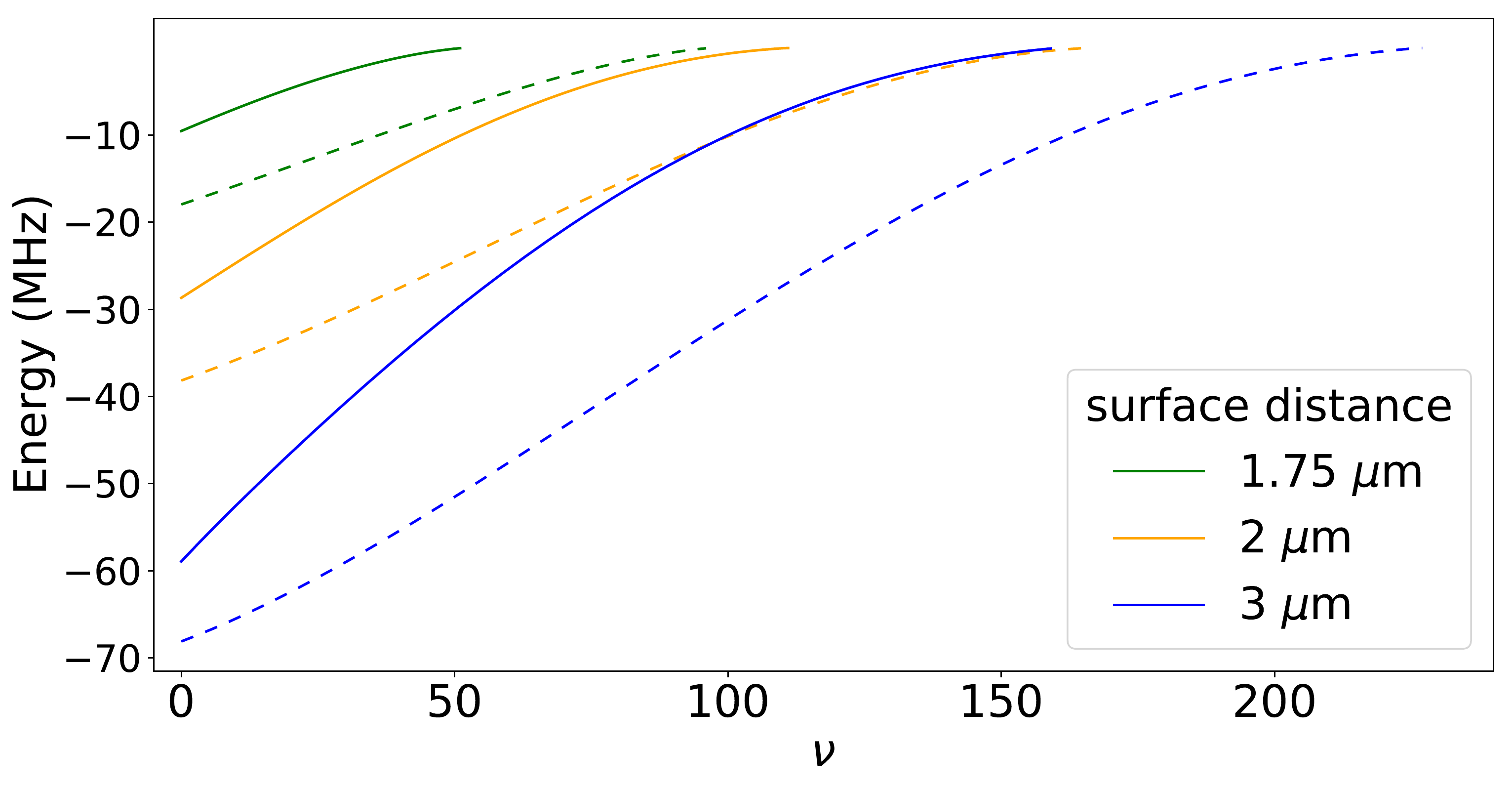}
\caption{Energies of the macrodimer bound states versus vibrational quantum 
number $\nu$. Solid (dashed) lines represent states in the upper (lower) 
potential branch, respectively.}
\label{fig:energyspectra}
\end{figure}

Figure~\ref{fig:energyspectra} shows the energy spectra of both upper and lower 
branches for surface distances $d_s=1.75\,\mu$m (green), $d_s=2\,\mu$m (orange), 
and $d_s=3\,\mu$m (blue). As the atoms approach the surface, the avoided 
crossing becomes more pronounced and the potential wells flatten resulting in 
fewer bound states, i.e. lower $\nu_\text{max}$ for smaller $d_s$.  For even smaller
atom-surface distances, $d_s \lesssim 1.5\,\mu$m, both upper and lower branch potential well
vanish completely. Therefore, bound macrodimers can only be found in a narrow window 
of surface distance $1.5\,\mu\mathrm{m} \leq d_s \leq 4\,\mu$m.

\begin{figure}[ht]
\includegraphics[width=\columnwidth]
{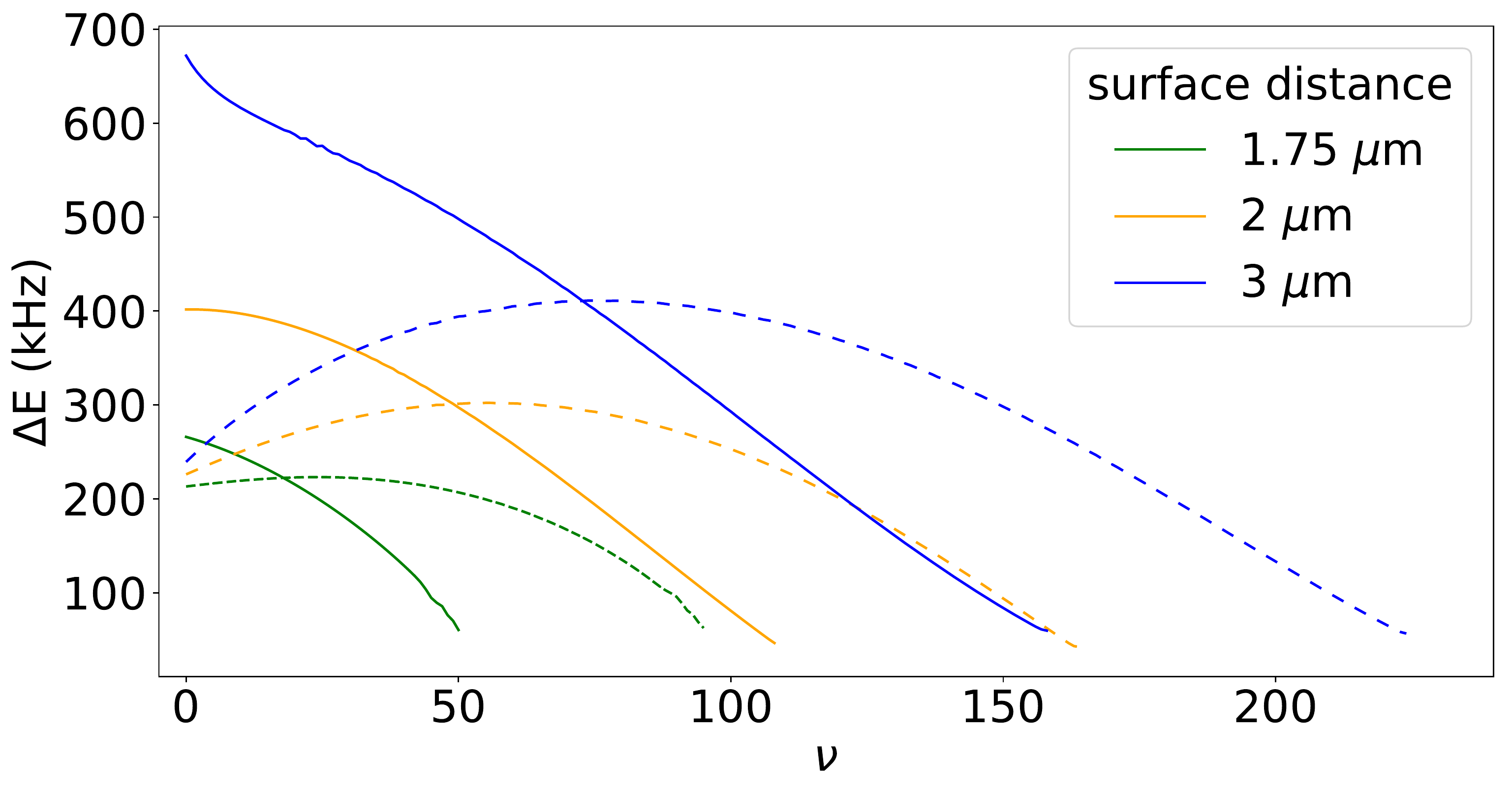}
\caption{Energy spacing $\Delta E_{\nu,\nu+1}$ between adjacent vibrational 
macrodimer states vs. vibrational quantum number $\nu$. Solid (dashed) lines 
represent states in the upper (lower) potential branch, respectively.}
\label{fig:energyspacing}
\end{figure}

While the energy of the deepest bound state is $E_\text{min} = E(\nu=0) \approx 
-59 (68)\,$MHz for the upper (lower) potential branch, states with large 
vibrational quantum number form a quasi-continuous regime  (see solid (dashed) 
lines in Fig.~\ref{fig:energyspectra}).
Even for deeply bound states, the energy difference $\Delta E_{\nu,\nu+1}$ between adjacent 
states is of the order of a few hundred kilohertz depending on the surface 
distance as shown in Fig.~\ref{fig:energyspacing}.
The peculiar shape of the lower potential branch with a very broad minimum 
produces a maximum $\Delta E_{\nu,\nu+1}$ at $\nu \gtrsim 30$ depending on $d_s$ 
(dashed lines in Fig.~\ref{fig:energyspacing}). At $d_s = 3\,\mu$m, the maximum 
energy spacing is $\Delta E_{\nu,\nu+1} \sim 410\,$kHz compared to $\Delta E_{\nu,\nu+1}\sim 240\,$kHz 
for the lowest vibrational states in the lower branch.
For smaller surface distance $d_s$, both the number of bound states 
$\nu_\text{max}$ and the energy spacing $\Delta E_{\nu,\nu+1}$ decreases, thus resulting in 
an effective continuum of states.

\begin{figure}
\includegraphics[width=\columnwidth]{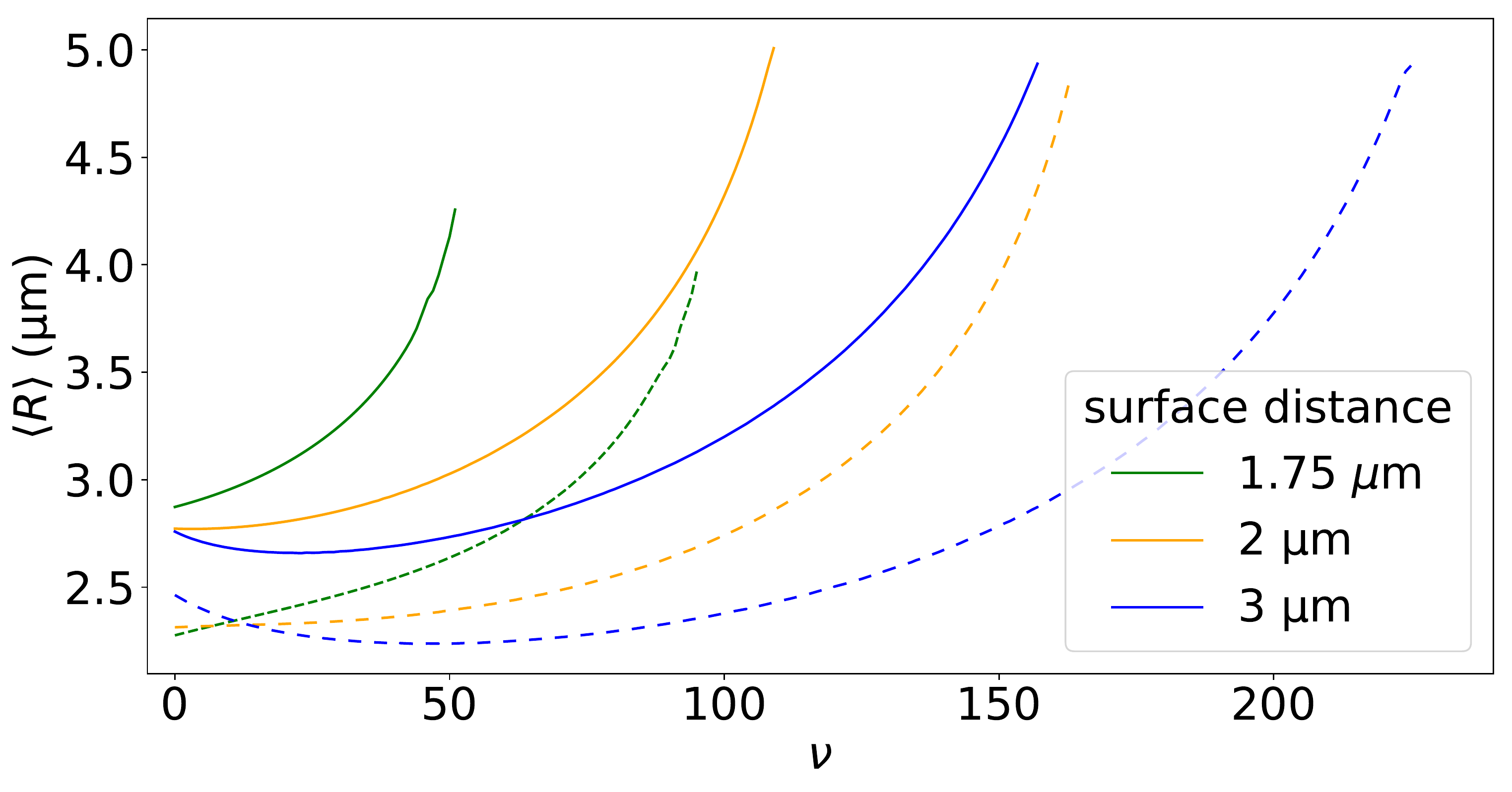}
\caption{Radial expectation value $\langle R\rangle$ versus vibrational quantum 
number $\nu$ for different surface distances. Upper (lower) potential branch 
represented by solid (dashed) lines. The minimum of some of the curves is a 
result of the anharmonic potential.}
\label{fig:radii}
\end{figure}

The mean interatomic distance of the macrodimer $\langle R \rangle = 
\langle\Psi |R|\Psi\rangle$ is shown for the upper (lower) potential branches
in solid (dashed) lines in Fig.~\ref{fig:radii}. For the lowest bound states, 
$\langle R\rangle (\nu = 0) \gtrsim 2.27 \,\mu$m depending on potential branch and 
surface distance. Because of the potential's anharmonicity, the radial 
expectation value decreases with increasing $\nu$ for $d_s > 2\,\mu$m and 
shows a pronounced minimum as a function of vibrational quantum number.
This behaviour can be explained by investigating the pair potentials in detail.
Figure~\ref{fig:potentials_and_wf} shows the anharmonic shape of both potential 
branches. The depicted wave functions suggest smaller radial expectation values 
of the dimer for $\nu = 20$ than for $\nu = 0$. The highest macrodimer 
states reach bond lengths of $\langle R\rangle_\text{max} \approx 5\,\mu$m.

\section{Discussion}
\label{sec:conclusions}

We have investigated Rydberg atom pair potentials in the presence of a 
perfectly reflecting surface. We have shown that the surface's influence on 
both single-atom energy levels 
and interatomic interaction leads to the creation of avoided crossings in the 
pair potential. Due to these avoided crossings, potential wells in the pair 
potential appear which we associate with long-range Rydberg macrodimer molecules 
having binding energies of upto $\sim 70\,$MHz and a bond length in the $\mu$m 
range. As the lifetime of a Rydberg state decreases in the proximity of 
a good conductor, the macrodimer's classical period of rotation is longer 
than the atomic lifetime, enabling us to safely neglect rotational states 
which is consistent with previous studies.

The maximum energy shift from the unperturbed pair state at the minimum 
of a macrodimer potential well combined with the contribution from the Casimir--Polder 
interaction for $R > 2\,\mu$m is $\Delta E_\text{max} \lesssim 170\,$MHz. 
This allows the selected excitation of atoms at a given surface distance 
\cite{Tong2004}. For example, the Casimir--Polder induced energy difference 
between the $|51s_{1/2};53s_{1/2}\rangle$ asymptote at $d_s = 2\,\mu$m and 
asymptotes at $d_s = 1.75\, (2.25)\,\mu$m is $\Delta E \sim 50\, (30)\,$MHz for 
large interatomic separations.

It is well known from the theory of atomic gases that the combination of a 
short-range repulsive potential and a longer-range attractive potential 
may lead to the formation of molecular crystals \cite{Ashcroft1976}.
As a simple approximation to the correct pair potential taken at $d_s = 
2\,\mu$m, we choose a Lennard-Jones (12,6) potential of the form 
\begin{equation}
 V(R) = 4\epsilon \left[\left(\frac{\sigma}{R}\right)^{12} - \left(\frac{\sigma}{R}\right)^6\right].
\end{equation}
This form of the potential yields a reasonably good approximation for large 
separations associated with the attractive van der Waals $R^{-6}$ part, 
whereas the short-range repulsive behavior contained in the 
$R^{-12}$ term reproduces the exact potential less accurately.
Nonetheless, the simple form of the Lennard-Jones potential allows one to fit 
the parameters $\epsilon, \sigma$ and analytically extract the cohesive energy 
of the corresponding two-dimensional crystal.

For the given parameters, we arrive at an equilibrium distance $R_\text{eq} 
\approx 2.75\,\mu$m and equilibrium cohesive energy of $E_\text{eq} \approx 
-93\,$MHz.
While the equilibrium distance is of the order of the mean macrodimer bond 
length calculated previously, the cohesive energy is enhanced by a factor of 3.2 
compared to the ground-state energy of the macrodimer at $d_s = 2\,\mu$m,
$E_\text{eq} = 3.2E_\text{min}$. These results suggest the possibility of 
crystal formation in an atomic ensemble close to a surface.
Combined with recent studies on macrodimer excitation in an optical lattice 
\cite{Hollerith2019}, this could yield a stable two-dimensional Rydberg atomic 
crystal even with external electromagnetic fields turned off.

In a possible experimental realisation of this idea, one important challenge 
will be unintentional adsorption of atoms to the surface 
\cite{Tauschinsky2010,Epple2014,Langbecker2017}.
For some $\sim10^8$ atoms adsorbed to the surface, the electric field generated 
by the adsorbates reaches a strength of $1\,$V/cm at a distance of 
$30\,\mu$m away from a copper surface \cite{Hattermann2012}.
This is substantially larger than the electric fields used to engineer 
macrodimers in previous studies that are typically in the range of $0.1\,$V/cm 
\cite{Schwettmann2007}. The effect described here will therefore be observable 
in situations with relatively few adsorbed atoms and, generically, the effects 
of (stray) electric fields and surface-induced interactions will have to be 
taken into account together.

\acknowledgments
We thank S. Weber for advice and assistance regarding the \textsc{pairinteraction} software 
and H. Stolz for fruitful discussions regarding the pair potential.
This work was partially supported by the Priority Programme 1929 "Giant Interactions in Rydberg Systems`` 
funded by the Deutsche Forschungsgemeinschaft and the Landesgraduiertenf\"orderung Mecklenburg--Vorpommern.

\FloatBarrier

\bibliography{./mendeley2019.bib}
\bibliographystyle{apsrev4-1} 

\end{document}